\documentstyle[prl,preprint,aps,epsf]{revtex}
\begin{document}
\draft
\preprint{GTP-97-05}
\title{ The Second Virial Coeffecient of Spin-1/2 Interacting Anyon System}
\author
{$^{a}$Sahng-kyoon Yoo and $^{b}$D. K. Park}
\address
{$^{a}$Department of Physics, Seonam University, Namwon 590-711, Korea \\
 $^{b}$Department of Physics, Kyungnam University, Masan 631-701, Korea}
\date{\today}
\maketitle
\begin{abstract}
Evaluating the propagator by the usual time-sliced manner, we use it to 
compute
the second virial coefficient of an anyon gas interacting through the 
repulsive potential of the form $g/r^2 (g > 0)$.
All the cusps for the unpolarized spin-1/2 as well as
spinless cases disappear in the $\omega \rightarrow 0$ limit, where $\omega$
is a frequency of harmonic oscillator which is introduced as a 
regularization method. As $g$ approaches to zero, 
the result reduces
to the noninteracting hard-core limit.
\end{abstract}

Since the anyon whose statistics interpolates between boson and fermion
at two-dimensions\cite{leinaas77,wilczek82,wilczek90} was introduced,
until recently the main focus has been on the free anyon gas, i.e., 
noninteraction apart from statistical interacton of Aharonov-Bohm type.
In order to investigate the statistical properties of a free anyon gas
the thermodynamic quantities such as the second virial coefficient as a
function of statistical parameter $\alpha$ has been calculated for both
spinless\cite{arovas85,comtet89} and spin-1/2 cases\cite{blum90}. 
The second virial coefficient of the spinless case shows the periodic 
dependence on $\alpha$ and nonanalytic behavior at bose points. 
However, for spin-1/2 case the discontinuities appear at bose points and 
periodicity is also removed.
This difference comes from the fact that the introduction
of spin allows the irregular wavefunction at origin. Even if
no irregular solution is assumed in the spin-1/2 case, 
the cusps exist at all integer points.
Recently we calculated the second virial coefficient for spinless and
spin-1/2 free anyon gases\cite{yoo97-1} for the  various values of self-adjoint 
extension\cite{capri85} parameter.
The result for spin-1/2 case exhibits a completely
different cusp and discontinuity structure from Ref. \cite{blum90}, due to
the different condition for the occurrence of the irregular wavefunction
at origin.

\indent Loss and Fu\cite{loss91} studied the interacting anyon gas with a
repulsive potential of the form $g/r^2$ ($g > 0$), using the similar
regularization procedure to that used in Ref. \cite{arovas85}. 
They chose 
$1/r^2$-potential, because it does not remove the scale invariance of theory
and the path-integral solution is simply obtained.
Furthermore, the probability of the overlap of two particles is always zero.
This property is also valid for spin-1/2 system with the same 
two-particle interaction\cite{park94}. They showed
that this simple interaction makes the cusps at bose points smooth for
spinless case.

\indent In this paper, we will compute the second virial coefficient 
of the spin-1/2 anyon gas 
interacting through this repulsive potential  
using the harmonic oscillator regularization. It is found that the cusps at
both boson and fermion points of the second virial coefficient calculated
under the condition that no irregular solution is assumed become
smooth as in the case of Loss and Fu\cite{loss91}. As $g \rightarrow 0$, 
the nonanalytic behavior of Blum {\it et al.}\cite{blum90} is reproduced.

\indent We begin with the kernel for the
anyon system with $g/r^2$ and harmonic oscillator interactions 
\begin{equation}
K [ {\bf r_f}, {\bf r_i}; T] = \int D{\bf r} e^{ i \int_{0}^{T}
dt L({\bf r}, \dot{\bf r}, t) },
\end{equation}
where
\begin{equation}
L( {\bf r}, \dot{\bf r}, t) = \frac{M}{2} \dot{\bf r}^2 - \alpha \dot{\theta}
- \frac{g}{r^2} - \frac{M}{2} \omega^2 {\bf r}^2
\end{equation}
is the Lagrangian of the system. Following the similar procedure to 
Ref. \cite{peak69}, one can obtain the euclidean kernel as
\begin{eqnarray}
G [{\bf r}_f, {\bf r}_i; \tau ] & = & \sum_{m = - \infty}^{\infty}
e^{im (\theta_f - \theta_i)} G_m [ r_f , r_i ; \tau ], \nonumber \\
G_m [r_f , r_i ; \tau ] & = & \frac{M \omega}{2 \pi \sinh \omega \tau}
\exp \left[ - \frac{M \omega}{2} \frac{\cosh \omega \tau}{\sinh \omega \tau}
(r_i^2 + r_f^2 ) \right] \\
& \times & I_{ \sqrt{(m + \alpha)^2 + 2 gM}} 
\left( \frac{M \omega r_i r_f}{\sinh \omega \tau} \right) \nonumber
\end{eqnarray}
where $I_{\nu} (x)$ is the modified Bessel function and $\tau = iT$.
Then we perform the Laplace transform to obtain the energy-dependent  
Green's function:
\begin{eqnarray}
\hat{G} [{\bf r}_f, {\bf r}_i; E ] & = & \sum_{m = - \infty}^{\infty}
e^{im (\theta_f - \theta_i)} \hat{G}_m [ r_f , r_i ; E ], \nonumber \\
\hat{G}_m [r_f, r_i; E] & = & \frac{1}{2 \pi \omega r_i r_f}
\frac{\Gamma \left( [ 1+\sqrt{(m+ \alpha)^2 + 2gM}+ E/\omega] /2 \right)}
{\Gamma \left( 1+ \sqrt{(m+ \alpha)^2 + 2gM} \right) } \nonumber \\
& \times & W_{- \frac{E}{2 \omega}, \sqrt{(m+ \alpha)^2 + 2gM}}
\left( M \omega [ \mbox{Max} (r_i ,r_f )]^2 \right) \\
& \times & M_{- \frac{E}{2 \omega}, \sqrt{(m+ \alpha)^2 + 2gM}}
\left( M \omega [ \mbox{Min} (r_i ,r_f )]^2 \right) \nonumber
\end{eqnarray}
where $W_{\kappa, \mu} (x)$ and $M_{\kappa, \mu} (x)$ are the usual Whittaker's 
functions, and $\mbox{Max} (x,y)$ and $\mbox{Min} (x,y)$ are 
the maximum and mininum values of $x$ and $y$, respectively. 
From the poles of the Green's function, the bound state spectrum of the system 
is straightforwardly obtained:
\begin{equation}
E_{n,m} = \left( 2n + 1 + \sqrt{(m+ \alpha)^2 + 2gM} \right) \omega.
\label{bound}
\end{equation}
The plot of $E_{0,0}$ at $gM = 1$ is shown in figure \ref{fig1}.
The cusps that happened in the absence of $1/r^2$ potential disappear and
become smooth. Therefore, by the introduction of repulsive potential, 
we expect that the nonanalytic dependence on $\alpha$ in various 
thermodynamic quantities would be suppressed.

\indent Now, we calculate the second virial coefficient $B_2$ of this
system. The two-particle partition function $Z_2$ is given by
\begin{eqnarray}
Z_2 & \equiv & \mbox{Tr} \exp ( - \beta H_2 ) \nonumber \\
& = & 2 A \lambda_T^2 Z_{rel},
\end{eqnarray}
where $H_2$ is the two-particle Hamiltonian, $\beta= 1/ kT$, $A$ is the area
of the system, $\lambda_T = (2 \pi / kTM)^{1/2}$ is the thermal de Broglie
wavelength, and $Z_{rel}$ is the partition function in relative coordinates.
The second virial coefficient then is
\begin{eqnarray}
B_2 (\alpha, T) & = & \frac{A}{2} - 2 \lambda_T^2 Z_{rel} \nonumber \\
& = & \frac{A}{2} - 2 \lambda_T^2 \sum_{n,m} e^{- \beta E_{n,m}},
\end{eqnarray}
where $E_{n,m}$ is given in Eq. (\ref{bound}) and $M$ is replaced by $2M$.
The summation over even (odd) $m$'s corresponds to the boson (fermion) 
statistics. At first performing the summation over $n$, we obtain
\begin{equation}
B_2 (\alpha , T) = \frac{A}{2} - \frac{ \lambda_T^2}{\sinh \beta \omega}
\sum_{m} e^{- \beta \sqrt{(m+ \alpha)^2 + gM} \omega}.
\end{equation}

\indent Consider the spinless case by summing only over even $m$'s. In order 
to regularize the infinite area, we calculate $B_2 (\alpha , T) - B_2 
(\alpha =0, T)$ :
\begin{equation}
B_2 (\alpha , T) - B_2 (0, T) = \frac{\lambda_T^2}{\sinh \beta \omega}
\sum_{m= even} \left[ e^{- \beta \sqrt{m^2 + gM} \omega} -
e^{- \beta \sqrt{(m+ \alpha)^2 + gM} \omega} \right].
\end{equation}
The result in the $\omega \rightarrow 0$ limit is just that of Ref.
\cite{loss91}.

\indent Next, consider the unpoliarized spin-1/2 anyon case. This can be
done by averaging over four possible spin states:
\begin{eqnarray}
B_2 (\alpha , T) - \bar{B}_2 (0, T) & = & 
\frac{\lambda_T^2}{4 \sinh \beta \omega}
\Bigg\{ 3 \sum_{m= odd} \left[ e^{- \beta \sqrt{m^2 + gM} \omega} -
e^{- \beta \sqrt{(m+ \alpha)^2 + gM} \omega} \right] \nonumber \\
& + & \sum_{m= even} \left[ e^{- \beta \sqrt{m^2 + gM} \omega} -
e^{- \beta \sqrt{(m+ \alpha)^2 + gM} \omega} \right] \Bigg\},
\end{eqnarray}
where $\bar{B}_2 (0, T)$ is the averaged $B_2 (0, T)$ which cannot be
determined but has no $\alpha$-dependence. We show $\omega \rightarrow 0$ limit
of $B_2 (\alpha , T) - 
\bar{B}_2 (0, T)$ as a function of $\alpha$ for $g=0, 0.05, 0.1$ and 1 in
figure \ref{fig2}. When $g > 0$,
the second virial coefficient has no cusps for all $\alpha$ as expected.
As $g \rightarrow 0$, the previous result\cite{blum90} is reproduced:
$| \alpha | - 2 \alpha^2$ for boson point and $3 |\alpha | - 2 \alpha^2$
for fermion point\cite{comment}. As a result, the repulsive interaction
removes all the cusps for spin-1/2 case as well as the spinless one.

\indent Even though $1/r^2$-potential is adopted to study the 
interacting anyons due to its simplicity, more realistic interaction
between anyons should be introduced in order to apply to real physical
systems. If we think anyons as the particles both carrying magnetic flux and
electrical charge, the consideration of Coulomb interaction arises naturally.
We have already calculated the kernal and bound states for 
Aharonov-Bohm-Coulomb system incorporating the self-adjoint extension 
method into the Green's function formalism appropriately\cite{park97}. 
Though the simple harmonic oscillator regularization
seems to be impossible because of difficulty in getting the path-integral 
solution for 
Aharonov-Bohm-Coulomb plus harmonic oscillator system, the second virial
coefficient may be obtained from the appropriate phase shift method in
scattering theory. This problem is now under study.

\indent In conclusion, we find the path-integral kernel for the
interacting spin-1/2 anyons with repulsive potential and harmonic oscillator,
and calculate the second virial coefficient using the partition
function obtained by summing the harmonic oscillator bound states.
For unpolarized spin-1/2 anyons, all the cusps at both boson and fermion 
points disappear, as in the spinless case. The nonanalytic behavior with
$\alpha$ is reproduced when $g \rightarrow 0$.

\begin{figure}
\caption{The bound state energy as a function of $\alpha$ when $n, ~m = 0$.
  Solid line : $gM =1$ case. Dashed line : $g = 0$ case in the presence of 
  irregular solution. Dotted line : $g = 0$ case in the absence of 
  irregular solution.}
\label{fig1}
\end{figure}

\begin{figure}
\caption{$[B_2 (\alpha, T) - \bar{B}_2 (0, T)] / \lambda_T^2$ as a function
   of $\alpha$ at various $gM$ values. Thick solid line : $gM = 0$. 
   Dotted line : $gM = 0.05$. Short-dotted line : $gM = 0.1$. Thin solid
   line : $gM = 1$.}
\label{fig2}
\end{figure}

\newpage
\epsfysize=20cm \epsfbox{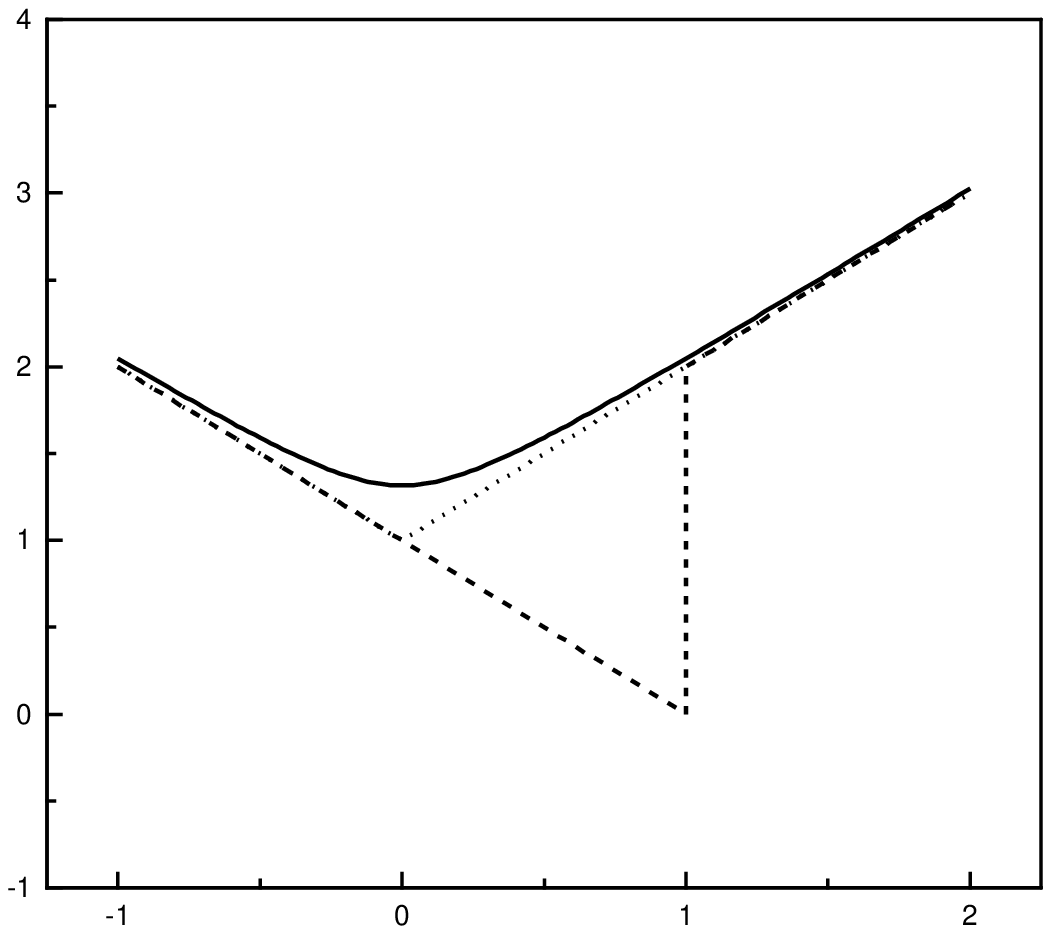}

\newpage
\epsfysize=20cm \epsfbox{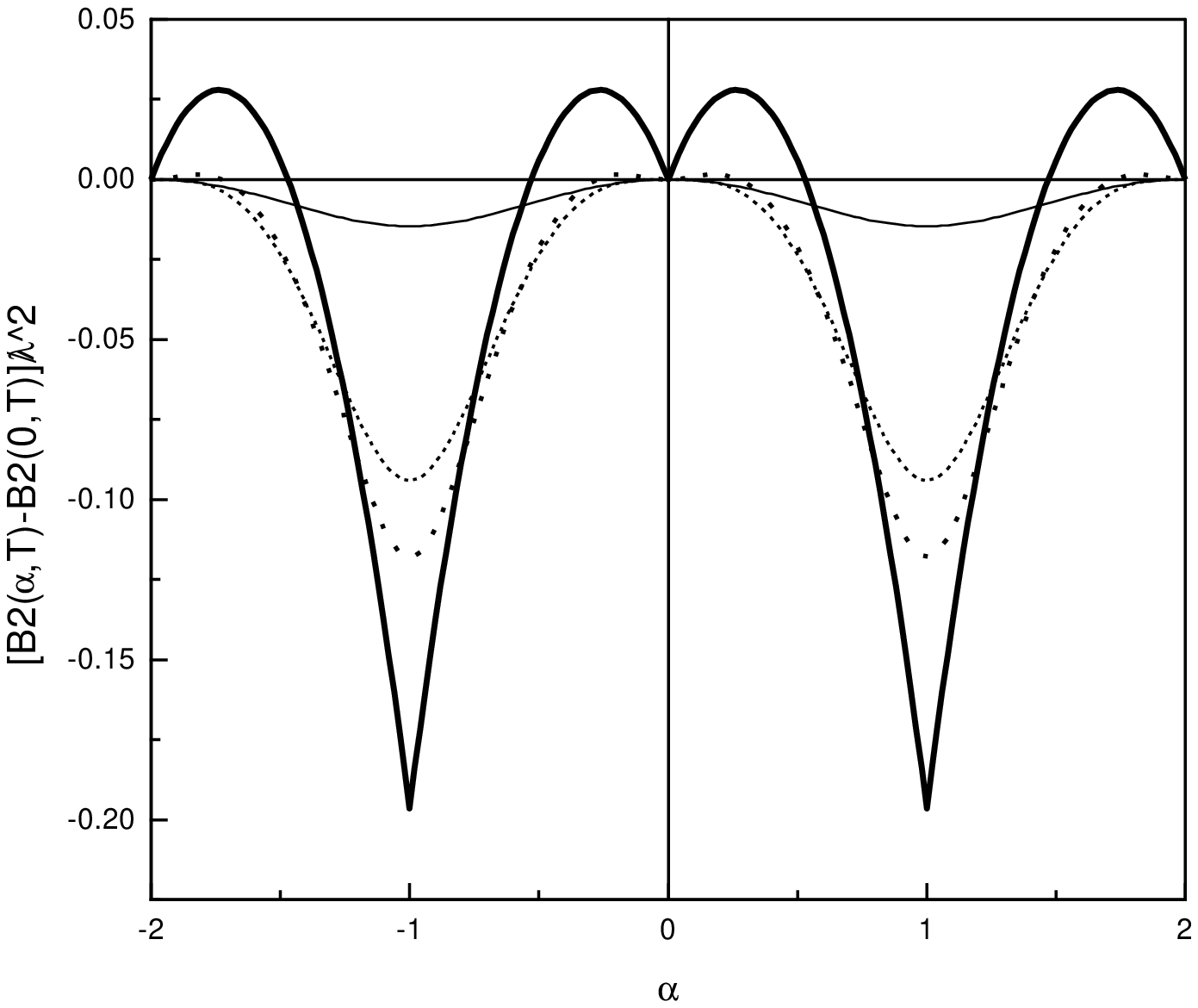}


\begin{references}
\bibitem{comment} Of course, the order of limits $\alpha, g \rightarrow 0$
   is also crucial in this case. See the detailed discussion in 
   Ref. \protect\cite{loss91} for this problem.
\end{references}
\end{document}